\documentstyle[psfig]{elsart}

\begin{document}
\maketitle

\begin{frontmatter}

\title{\sc A Parallel Tree code for large Nbody simulation: dynamic load balance and data distribution on CRAY T3D system} 

\author[oact]{U. BECCIANI\thanksref{cnr}},
\author[cray]{R. ANSALONI},
\author[oact]{V. ANTONUCCIO-DELOGU\thanksref{cnr}}, 
\author[cineca]{G. ERBACCI},
\author[oact]{M. GAMBERA\thanksref{cnr}}, and
\author[iact]{A. PAGLIARO\thanksref{cnr}}

\address[oact]{Osservatorio Astrofisico di Catania,
Citt\`{a} Universitaria, Viale A. Doria, 6 --
 I-95125 Catania - Italy}
\address[cray]{Silicon Graphics S.p.A. St.6 Pal.N3 Milanofiori
I-20089 Rozzano (MI) - Italy}
\address[cineca]{Cineca, Via Magnanelli,6/3 I-40033 Casalecchio di Reno (BO)- Italy}
\address[iact]{Istituto di Astronomia, Universit\`{a} di Catania,
Citt\`{a} Universitaria, Viale A. Doria, 6 --
 I-95125 Catania - Italy}
\thanks[cnr]{Also: CNR-GNA, Unit\`{a} di Ricerca di Catania}

\begin{abstract}
\noindent N-body algorithms for long-range unscreened interactions like gravity 
belong to a class of highly irregular problems whose optimal solution is a 
challenging task for present-day massively parallel computers.
In this paper we describe a strategy for optimal memory and work distribution 
which we have applied to our parallel implementation of the Barnes \& Hut (1986)
recursive tree scheme on a Cray T3D using the CRAFT programming environment. 
We have performed a series of tests to find an {\it optimal data distribution} 
in the T3D  memory, and to identify a strategy for the {\it Dynamic Load 
Balance} in order to obtain good performances when running large simulations 
(more than 10 million particles).
The results of tests show that the step duration depends on two main factors: 
the data locality and the T3D network contention. Increasing data locality 
we are able to minimize the step duration if the closest bodies (direct 
interaction) tend to be located in the same PE local memory (contiguous 
block subdivison, high granularity), whereas the tree properties have a 
fine grain distribution. 
In a very large simulation, due to network contention, an unbalanced load  
arises. To remedy this we have devised an automatic work redistribution 
mechanism which provided a good Dynamic Load Balance at the price of 
an insignificant overhead.
\end{abstract}

\end{frontmatter}

\section{\bf Introduction}

N-body simulations are one of the most important tools by which contemporary theoretical
cosmologists try to investigate
the evolution of the Large Scale Structure of the Universe. Due to the long-range
nature of the gravitational force, the number of particles required to reach a
significant mass resolution is a few orders of magnitude larger than those allowed even 
by present-day state-of-the-art massively parallel supercomputers (hereafter MPP).
Work- and data-sharing programming techniques are customary tools exploited in 
the development of many parallel implementations of the N-body problem \cite{xg95} \cite{rdh97} \cite{js97} . 
The most popular algorithms are generally based on grid
methods like the $P^3M$. The main problem with this method lies in the fact that the
grid has typically a fixed mesh size, while the cosmological problem is inherently
highly irregular, because a highly clustered Large-Scale Structure develops
starting from a nearly homogeneous initial
distribution.  On the other hand the Barnes \& Hut (1986) oct-tree
recursive method  is inherently adaptive, and allows one to achieve a higher
mass resolution than grid-based methods when clusters of galaxies form.
Because of these features, however, the computational problem can easily run into
unbalance causing a performance degradation. For this reason, we have 
undertaken a study of the 
optimal work- and data-sharing distribution for our parallel treecode.\\
Our Work- and Data-Sharing Parallel Tree-code (hereafter WDSH-PTc)  
is based on this algorithm tree scheme, which we have modified to run on a
shared-memory MPPs \cite{bap96} \cite{ab94}.
We have adopted the Cray Research Corporation CRAFT environment \cite{cr94}
to share  
work and data among the PEs involved in the run.\\
To optimize the performances of the WDSH-PTc in order to run simulations 
with a very high number of particles, we have carried out a study on the
 optimal data distribution in the T3D global memory \cite{jb97}. 
The obtained results  allow us to determine an optimal strategy for the Dynamic Load 
Balance (DLB), that is the main purpose of this work. Generally all tests and 
measurements were carried out using a $\theta$ value equal to 0.8, considering this 
value as the optimal value for the simulations that we would like to run.

\section{\bf The Barnes-Hut based WDSH-PTc}

The 
Barnes-Hut Tree algorithm is a $NlogN$ procedure to compute the gravitational force 
through a hierarchical subdivision of the
computational domain in a set of cubic nested regions. 
The evolution of the system involves the advancement of the trajectories of all 
particles, and this is carried out through a discrete integration of the trajectories
of each particle. At each timestep the force
and the trajectory of each body are updated. In practice one does not adopt the ``nude''
gravitational potential, in order to avoid the formation of binary systems, but a
potential smoothed on a sufficiently small scale.
A more detailed discussion on the BH tree method can be found in \cite{bh86}.\\
For our purposes, we can distinguish three main phases in 
each timestep. The first is the 
TREE\_FORMATION (hereafter TF) phase where the tree structure is built
starting from the whole computational domain included in a cubic region,
that is the "root-cell" (i.e. the zero level) of the tree. Using the ORB 
(Orthogonal Recursive Bisection) technique, at the begining
the root-cell is subdivided into 8 cubic nested regions (subcells) each 
including a portion of the computational domain. This subdivision creates 
the second level of the tree. The ORB technique is then recursively applied to each new
subcell, so that 
several levels of the tree are formed until all the final cells contain at most one body.
 Tree cells containing more than
one body are called "internal cells" (icells), and those containg 
only one body are called "terminal cells" (fcells).  For each icell the total mass, the 
center of mass and the quadrupole moment are computed.\\
The second phase is the FORCE\_COMPUTE\ (hereafter FC), during which  the forces 
acting on each body of the system are computed. In the 
TREE\_INSPECTION (hereafter TI) subphase, for each body one makes an "interaction list" containing 
pointers to cells with which the particle will interact, formed according to the 
following criteria. Starting from the root-cell the $C/d$ ratio is compared
with a 
threshold parameter $\theta$ (the opening angle parameter), "C"
  being the cell-size and "d" the
 distance  between the particle and the cell center of mass.\\   
If the $C/d$ ratio is smaller than $\theta$, the cell is considered as a single 
region that contributes a force component to the particle, and its subcells are not
further investigated. Otherwise one checks all the
subcells using the same criterion, until one arrives at cells containing only one particle. At the 
end,  all the tree cells satisfying the criterion form the "body Interaction List" for the given
body.\\
In general, using  the value $\theta$ = 1, the acceleration approximation has
an error of about 1\%. Typical values of $\theta$ for cosmological simulations are 
in the range 0.5 - 1.2: the larger the $\theta$ the smaller the length
of the interaction list. The average length of the "Interaction List" is 
proportional to $logN$, so that the total computation complexity in a serial 
code based on the Barnes-Hut tree code scales as $O(NlogN)$. 
After the TI subphase, using the interaction list, the 
ACC\_COMPONENTS (hereafter AC)  subphase is executed and the particle 
acceleration is computed.\\
\begin{figure}
\psfig{file=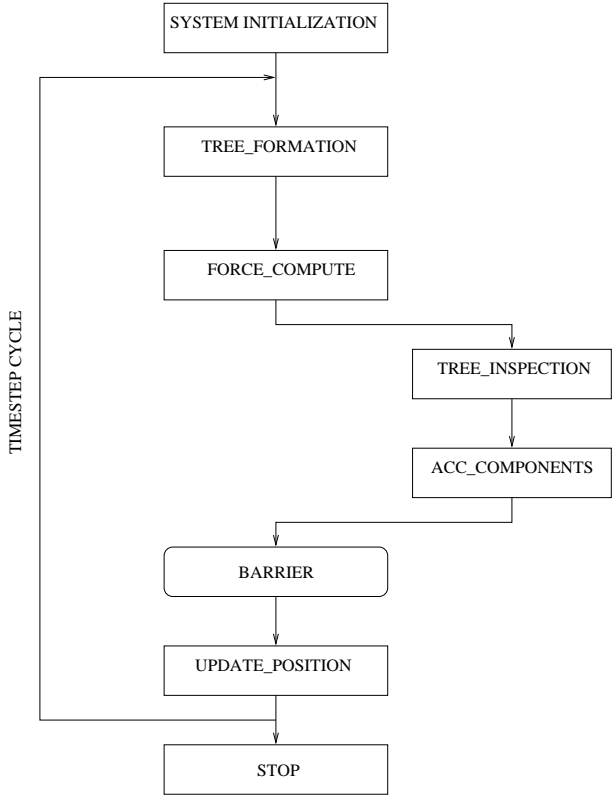,width=8cm}
\caption[h]{WDSH-PTc block diagram}
\end{figure}
At the end of the FC phase there is a synchronization point followed by
the UPDATE\_POSITION phase, the last phase of the timestep, when 
the bodies positions are updated (Fig. 1).

\section{\bf The WDSH-PTc  parallelism level}

The tests carried out on the original serial N-body code, kindly provided to us by Dr. L.
Hernquist, confirm the results shown by Salmon (1990) concerning the complexity. In
particular the total time spent in the TF phase ranges from 7\% to 
10\% of the overall time according to the particles initial condition; i.e., uniform 
or clustered distributions. Whereas the TI ranges from 65\% to 70\%,
the AC subphase last for the remaining 20\% - 25\% of the total timestep.
 The WDSH-PTc 
uses the work-sharing technique in order that all PEs cooperate in the TF 
phase, during which the tree is formed and the cell properties are computed, and the 
synchronization is implicit \cite{bap96}.\\
During the FC phase each PE computes the  acceleration components for each 
body in asynchronous mode and only at the end of 
the phase an explicit barrier statement is set.\\
The WDSH-PTc parallelism level reached is very high, more than 90\% of the work is 
performed in a parallel region as shown by Apprentice, a "performance analysis 
tool" designed for Cray 
MPP systems that allows one to investigate the effective performances reached by a code. 
With various pool configurations of PEs ranging from 16 to 128, our results show that the 
most time-consuming phases 
(TF, TI and AC) are 
executed in a parallel regime.\\
In the following sections we will discuss the performances obtained using 
different ways for data distribution in the memory and how a strategy of 
Dynamical Load Balancing (hereafter DLB) can be devised.

\section{\bf  WDSH-PTc T3D data distribution and performances}

Several strategies
to share data in the T3D PEs memory \cite{jb97} \cite{cr93} were investigated 
to find the best data distribution, in order to maximize the number of 
computed particles per second. 
We have considered the two main kinds of data, tree data (cells pointers and 
cells properties) and body data, and we have observed the code 
performances by varying the tree data and the body data 
distribution.\\
Tests were  carried out, fixing the constraint that each PE executes the 
FC phase only for all bodies residing in the local memory. 
A bodies data distribution ranging from contiguous blocks (coarse grain:
CDIR\$ SHARED POS(:BLOCK,:)) to a fine grain distribution (CDIR\$ SHARED 
POS(:BLOCK(1),:)) was adopted. We studied different tree data distributions 
ranging from assigning to contiguous blocks a number of cells equal to the expected 
number of internal cells (NTOTCELL), as  described in J. Salmon (1990) 
(coarse grain: CDIR\$ SHARED  POS\_CELL(BLOCK(:NTOTCELL/N\$PES),:)), to a simple fine 
grain distribution (CDIR\$ SHARED POS\_CELL(:BLOCK(1),:)).\\
All the tests were performed for two different set of initial conditions, namely 
uniform and clustered  distribution
having $2^{20}$ particles each. The tests were carried out using from 16 to 
128 PEs, and in 
Tab. 1  we report only the most significant results obtained 
with 128 PEs and using coarse grain and fine grain data distribution, although the 
same trend was obtained using 16 and 64 PEs.

\subsection{\it Tree data distribution considerations}

The data (i.e. particles and cells) distribution greatly affects the overall 
code performance and an accurate study has to be carried out
to obtain high gain from the MPP machines for this kind of code.
One possible approach was adopted by \cite{d96}, and is based on the so-called ``Locally
Essential Tree'' (LET) introduced by \cite{s90}, where each PE builds a ``local tree'' for
the particles assigned to it, and the force acting on each particle is then computed
after all the PEs have builded a LET assembling together the pointers to all the
cells from other PEs which do contribute (according to the above mentioned
$\theta$-criterion) to the force on all their particles. The main problem with this 
approach is that the memory requirements grow very quickly with the number of particles 
$Nbodies$. We have then preferred to keep a single tree shared among all the PEs and
to look for the optimal distribution of tree's cells and particles.\\
The tree data distribution greatly affects the overall code performance
and the distribution must be thoroughly carried out
to obtain high gain from the MPP machines for this kind of code. Our results
 show that the best tree data distribution is obtained using 
a block factor equal to 1, degenerate in the second dimension 
(CDIR\$ SHARED POS\_CELL(:BLOCK(1),:)). This is in accordance with
 what we expected, and in order to understand this point we notice that 
there are two aspects of the problem to be considered.\\
The first  is related to the cell inspection performed during the 
FC  phase. 
During the  TF phase, cells properties are 
determined level by level, starting from  level 0 (containing only 
the root cell) and descending down to deeper levels of depth L (each L level has 
$2^{L*3}$ cells). The cells are numbered 
progressively starting from the root cell. 
Considering that all the cells belonging to the first four low levels generally
include a large part of the domain, they are inspected by 
each PE  $Nbodies/N\$PES$ times during each timestep because that is the average\
 number of bodies that each PE has to treat. 
A fine grain tree data distribution involves that each PE have the same 
number of cells ($\pm 1$) and cells belonging to low levels are distributed 
over the PEs local memory. Using this kind of  tree data distribution, 
we obtain that for each PE the execution time of the FC phase is almost the same, because
all the PEs spend on average the same amount of time spent in the tree data access. 
Results in Tab .1 show that, a coarse grain tree data distribution 
increases the duration of the TF phase and the number of particles per
second, executed in a timestep, up to a factor of five. This is due to the network
contention to access to  "critical resources". In the case of a coarse
grain tree data distribution, all the cells belonging to the first
levels are located in the  first PE (or in the first two PEs), 
and all PEs access them at the same  time, during the FC phase.\\
Another aspect is related to the highly dynamical evolution of tree's properties.
 Each timestep produces a different tree, making it 
very difficult to  determine rules for an optimal tree data distribution which 
can minimize the access time on  the T3D toroidal architecture. Because of the
 overhead due to data redistribution during
a run, as well as the fact that the Block Factor power of two (imposed by CRAFT in 
the T3D) we deem it inconvenient to further examine this point.
So it is possible to conclude that, as shown  in Tab. 1, a fine grain tree data 
distribution should be used for
this kind of codes. 
  
\subsection{\it Bodies data distribution}

Bodies are labelled in such a way that close bodies have adjacent numbers,
and the properties are shared among the PEs  using the CRAFT directive  
CDIR\$ SHARED POS(:BLOCK(N),:)  N ranging between $1$ and $Nbodies/N\$PES$.\\
The fine grain bodies data distribution (bf) is obtained using a Block
factor $N=1$; i.e.,  bodies are shared among the PE but there is no  spatial 
relation in the body set residing in the same PE local memory. 
The medium grain bodies data distribution (bm) is obtained using a Block
factor $N=Nbodies/ 2*N\$PES$. Using this kind of distribution each PE 
has  two data block of  bodies properties residing in the local memory,
each block having a close bodies set.
At the end the  coarse grain bodies data distribution (bc) is obtained using
a Block factor $N=Nbodies/N\$PES$; i.e., each PE has one close data set block
of bodies residing in the local memory.
In any case, each PE executes the FC phase only for those bodies residing in
the local memory. The results reported in  Tab. 1 show that the 
best bodies data distribution, having the 
highest  code performance 
in terms of particles per second, is obtained using the Block factor 
N=Nbodies/N\$PES as expected. The most time-consuming subphases are the TI
 and the 
AC. During the first phase the  body interaction 
list is formed, containing some tree cells and close bodies.
Generally, tree cells are shared among all PEs, whereas bodies are 
residing in the same PE  or in the nearest PEs. This fact reduces the 
access time of the bodies properties included in the interaction list.
Therefore the
obtained results confirm that by distributing  the close  bodies in the same PE
 (coarse grain)  we obtain the best performance.\\
This effect depends on the order of the bodies so that nearest bodies have 
nearest numbering in the arrays containing bodies properties. If necessary,
 a sorting of the array can be performed, to obtain higher performances.

\section{\bf Dynamic Load Balance}

As stated above, the best choice is to have a fine grain tree data 
distribution and  a coarse grain bodies data distribution.
As emphasized by the Unbalance Factor in Tab. 1, it is very
important to adopt a strategy allowing to increase the load balance and 
obtain higher performances.
At the beginning \cite{bap96}, we adopted  a DLB technique 
based on the concet of ``PE executor'', i.e., 
the PE executing the FC phase for the body. The PE executor was re-assigned
at each time step in order to balance the load.\\
Although this usually brings some advantages, sometimes the overhead
due to the PE executor re-assignment may greatly reduce the usefulness of this scheme.
Here we present the results of a new DLB strategy, that allows us to avoid 
any large overhead, because no explicit control mechanism is necessary.
The total time spent in a parallel region $T_{tot}$, can be considered as the sum of the following terms\\

\begin{equation}
T_{tot}=T_{s}+ K T_{p}/p + T_{o}(p)		
\end{equation}

where $p$ is the number of processors executing the job,
 $T_{s}$ is the time spent in the serial portion of the code (a typical MASTER 
region), $T_{p}$ is the time spent by a single processor ($p=1$) to execute the parallel region,
$T_{o}(p)$ the overhead time due to the remote data access and to the synchronization 
points, and $K$ is a costant.\\

\begin{tabular} {|l|l|l|l|l|l|} \hline

	& PE\#	&p/sec	&FC phase	&T-step	&UF \\  \cline{1-6}
1Mun\_tf\_bf &	128 &	4129 &  230.05 &249.5 &4.22 \\ 
1Mcl\_tf\_bf &	128 &	3832 &	250.32 &	268.81 & 4.57 \\ 
1Mun\_tf\_bm &	128 &	3547 &	270.51 &	290.45 & 5.90 \\
1Mcl\_tf\_bm &	128 &	3308 &	291.63 &	312.26 & 6.32 \\ 	
1Mun\_tf\_bc &	128 &	4875 &	186.31 &	205.32 & 4.14 \\ 
1Mcl\_tf\_bc &	128 &	4490 &	203.37 &	222.72 & 4.38 \\ 
1Mun\_tc\_bc &	128 &	837  &	1051.93 &	1230.0 & 16.33 \\ 
1Mcl\_tc\_bc &	128 &	750 &	1173.24 &	1373.4 & 17.62 \\  \hline

\end{tabular}

\vspace{1cm}

\hspace {5cm} Tab .1 

\vspace{0.5cm}

\begin{tabular} {ll} 

Legend:& 1Mun - 1 million of particles in uniform initial conditions;\\
&	1Mcl - 1 million of particles in clustered initial conditions;\\
&	tf - tree fine grain data distribution \\
&	tc - tree coarse grain data distribution \\
&	bf - bodies fine grain data distribution \\
&	bm - bodies medium grain data distribution \\
&	bc - bodies coarse grain data distribution\\
\\
&	PE\#: 	PEs number \\
&	p/sec: 	particle per second \\
&	FC: 	total FC pashe duration in second \\
&	T-step:	timestep duration in second \\
&	UF: Unbalance Factor (variance of FC phase duration)\\
\\
\end{tabular}
\newpage

In the FC phase, there are no serial regions, so the $T_{p}$ term is
proportional to the length of the interaction list needed to compute the force acting 
on each body, the  interaction list average length  being $O(log(Nbodies))$.
Using a coarse grain subdivision, each PE has a block of close bodies in the 
local memory ($Np=Nbodies/N\$PES$); in a uniform distribution initial condition, the PEs having 
extreme numeration in the 
pool of available PEs, have residing bodies near the border of the computational domain.
Owing  to the lack of bodies besides the border line of the
computational domain (i.e. these bodies have a smaller interaction list than bodies in the
 center of the domain), the PEs having extreme numeration have a lower load at each timestep. 
This kind of effect may be enhanced, if a clustered initial condition is used.\\
\begin{figure}
\psfig{file=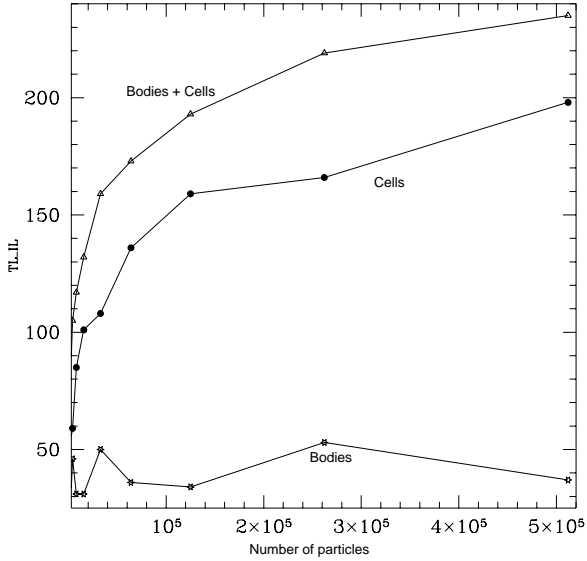,width=8cm}
\caption[h]{Interaction list with $\theta =0.8$}
\end{figure}
The $T_{o}(p)$ term depends principally on the latency time, on the bandwidth of the internal 
network, on the code synchronization point, and on the network contention.
When the number of PEs involved in the simulation increases the data dispersion on the
T3D torus increases.\\
In Fig. 2 we plot the total length of the interaction list, and the total number of cells
 and 
bodies forming the interaction list, using  $\theta=0.8$. These data were obtained making 
several tests 
with uniform and clustered initial conditions and the results are in accordance with plots 
reported in Salmon (1990). 
The number of internal cells included in the interaction list  ranges from
 $2/3$  to $4/5$ of its total length. The tree cells have a 
fine grain distribution as stated above, and then data access to these elements increases
with raising PE number which means the $T_{o}(p)$ term in (1) increases.\\
\begin{figure}
\psfig{file=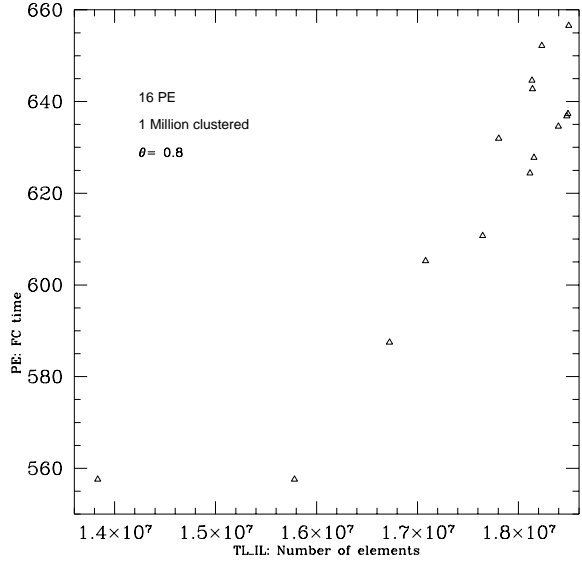,width=8cm}
\caption[h]{16 PE run}
\end{figure}
\begin{figure}
\psfig{file=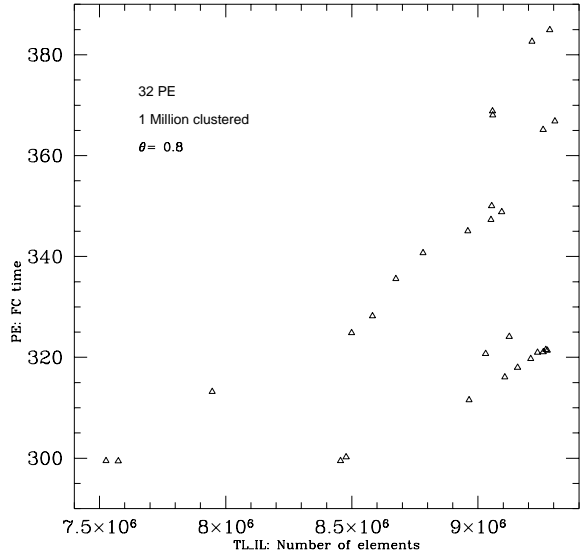,width=8cm}
\caption[h]{32 PE run}
\end{figure}

\begin{figure}
\psfig{file=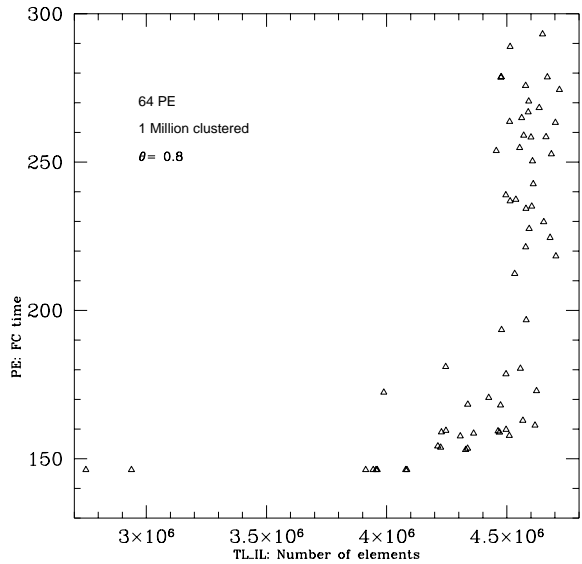,width=8cm}
\caption[h]{64 PE run}
\end{figure}

The Figs. 3,  4 and 5 report the total time ($T_{tot}$ in (1)) in seconds
 spent in the 
FC phase (T\_FC) by each PE, for different values of 
 the total length of the interaction list (TL\_IL), using 16, 32 and 64 PEs, respectively with 1 million of clustered particles $\theta=0.8$. 
TL\_IL value is the sum of all the
interaction lists obtained during the FC phase performed by the single PE.
Fig. 6 shows a similar result obtained by running a test with only 1 PE
on the T3D machine. In this case all bodies and tree cells properties
are located in the PE local memory. The results show a linear dependence
 between TL\_IL and  the $T_{p}$ term. A comparison between data reported in Fig.
6 and Figs. 3, 4 and 5 leads us to the following consideration. When 16 PEs are 
used (Fig. 3) a relationship 
between the T\_FC  and TL\_IL terms may be found. The overhead time $T_{o}(p)$, ranging from 60\% up to 70\% of 
the $T_{p}/p$ term in (1), can be calculated from
 the difference between data reported in Figs. 3 and 6. 
When the PE number increases the relationship is lost as shown Fig. 5.
The $T_{o}(p)$ increases
as the PE number increases, whereas the code performances increase when 
the load balance is optimized.\\
\begin{figure}
\psfig{file=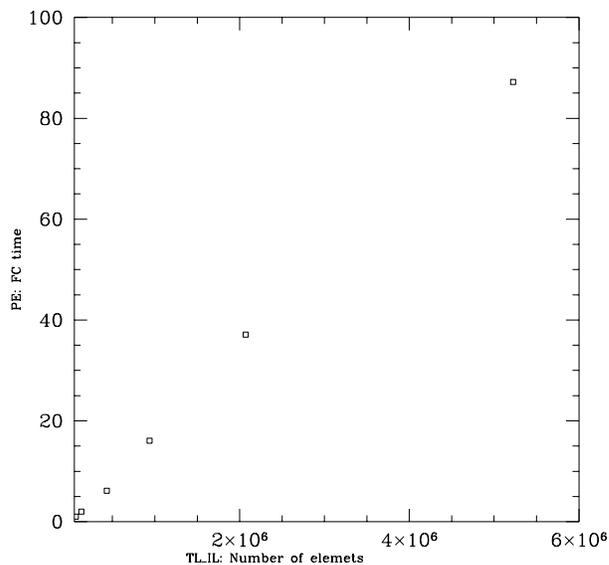,width=8cm}
\caption[h]{1 PE run}
\end{figure}
The adopted technique is to perform a load redistribution
among the PEs so that all PEs have the same load in the FC phase. 
We force each PE to execute this phase {\it only} for a fixed portion
 NB\_LP of the bodies residing in the
local memory.  The NB\_LP value is calculated as

\begin{equation}
NB\_LP=(Nbodies/N\$PES)*P\_LP		
\end{equation}

the P\_LP being a constant ranging from $0$ to $1$. The FC phase for all 
the remaining bodies 

\begin{equation}
Nfree=N\$PES*(Nbodies/N\$PES)*(1 - P\_LP)  		  
\end{equation}

is executed from all the PEs that have finished the FC
 phase for
the NB\_LP bodies. No correlation between the PE memory, where the
body properties are residing, and the PE, executing the FC phase for it, is found.\\
If P\_LP=1 all PEs execute the FC phase only for bodies residing in the  local memory. 
$Nfree$ is equal to 0
and the body locality is totally involved in the overall performances,
on the contrary if P\_LP=0, 
$Nfree=Nbodies$ (NB\_LP =  0),  the PEs execute only Nfree bodies and the locality is
 not taken into account.\\
A P\_LP value lower than 1 gives an automatic  dynamic  load
balance mechanism to the system; i.e., each PE works until all the Nfree bodies are computed.
On the other hand, if  P\_LP value is equal to 0, it gives the maximum load balance and the maximum
 degree of automatism in the FC phase.\\
Several tests were performed with P\_LP value ranging from 0 to 1 and PEs 
ranging 
from 16 to 128, using several  initial conditions  from 1000 up to 2 
million  particles uniform and clustered. Figs. 7-10 show the 
results obtained only with a high ($\geq 10^{5}$) number of particles, but the results 
obtained with smaller numbers of particles show the same trend.\\
Using the reported results and fixing the PE number, it is possible to
determine the  P\_LP value allowing the best 
code performances. In particular, note  
that for a simulation 
using  a high number of PEs (more than 32),  it is convenient  to fix the P\_LP value near
 to $0$,
that is  maximize the load balance among the PEs rather than the 
bodies locality. This is due to the  network contention
 of the system that becomes  relevant for the remote data access, mainly 
for the data access time of the tree cells.\\
\begin{figure}
\psfig{file=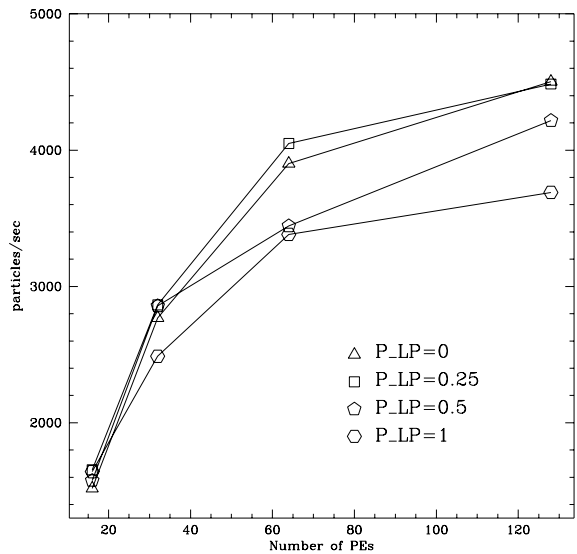,width=8cm}
\caption[h]{1 Million of particles: clustered configuration}
\end{figure}
On the other hand, using a lower PE number (from 16 to 32)  we observe a different effect. We
 have a lower 
data dispersion among the PEs, so the remote loads are fewer and have a shorter access time, thus
the code performance takes advantage of the data locality. From Figs. 7-10 it is possible
 to fix a P\_LP value combining the "load remote" effect and the "data 
locality" effect to maximize the computed  number of particles per second, thus improving the 
code performances.\\
\begin{figure}
\psfig{file=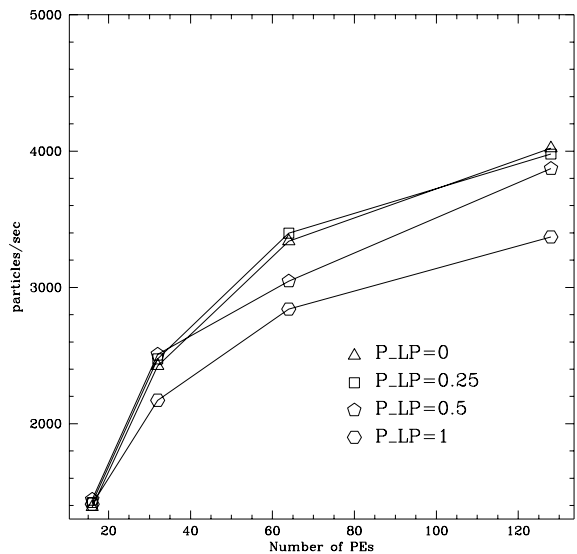,width=8cm}
\caption[h]{2 Million of particles: clustered configuration}
\end{figure}
The figures show that, fixing the PEs number and the particles number,  the 
same P\_LP value gives the best performance both in uniform and clustered conditions. 
This means that it is possible to fix a P\_LP value and this can be usefully adopted  
during all the running time: it is not necessary to recompute the P\_LP value to 
have good performances. Hence an automatic Load Balance mechanism
is found without  recalculating 
the P\_LP value and without introducing any overhead time to obtain a good 
Dynamic Load Balance.
\begin{figure}
\psfig{file=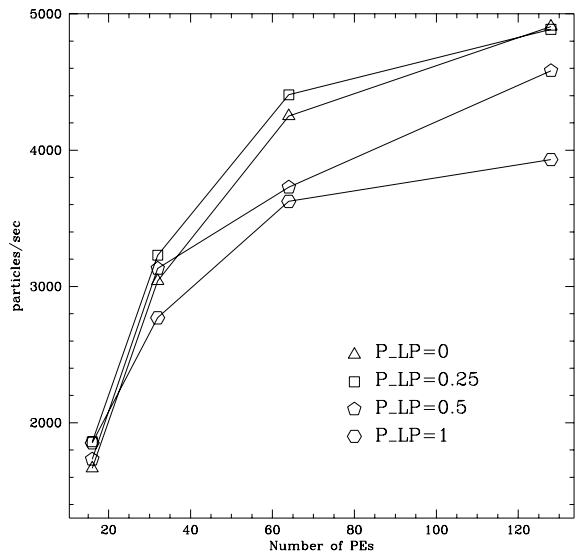,width=8cm}
\caption[h]{1 Million of particles: homogeneous configuration}
\end{figure}

\begin{figure}
\psfig{file=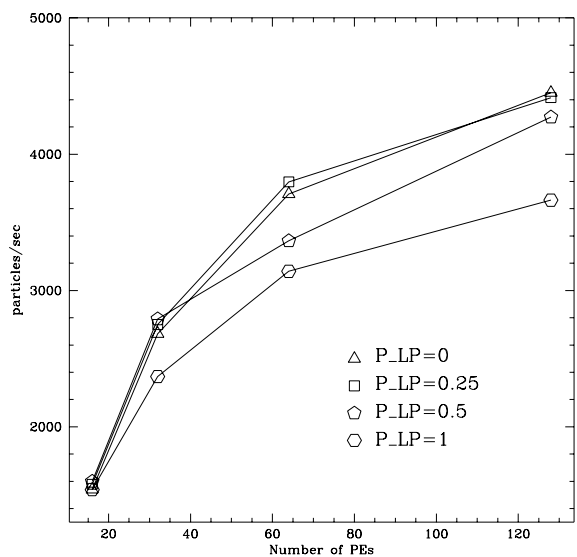,width=8cm}
\caption[h]{2 Million of particles: homogeneous configuration}
\end{figure}

\section{\bf Memory occupancy and performances: final considerations}

As stated in \cite{bap96} the  memory occupancy of this kind of N-body 
code is lower than the N-body   Local Essential Tree-based codes. The total memory 
occupancy of  WDSH-PTc, at present, allows to execute large simulation with more 
than 80 million of particles on  the CINECA T3E machine having 128 PEs  each with
16 Mword RAM. Moreover the total memory occupancy may be reduced in order to run larger
simulations with a little  degradation of code performances.\\
Fig. 11 shows the code speed (preliminary data) in number of particles per second, computed at each 
timestep, vs. the number of PEs,  using two 
$\theta$ values: $0.8$ and $1.2$, for a clustered configuration of
$10^{6}$ particles, using the periodic boundary conditions and adopting a 
 {\it grouping} strategy  \cite{b90},  \cite{gb}, \cite{bapg97} 
i.e. building  an interaction list for a group of bodies (included in a cell).\\
\begin{figure}
\psfig{file=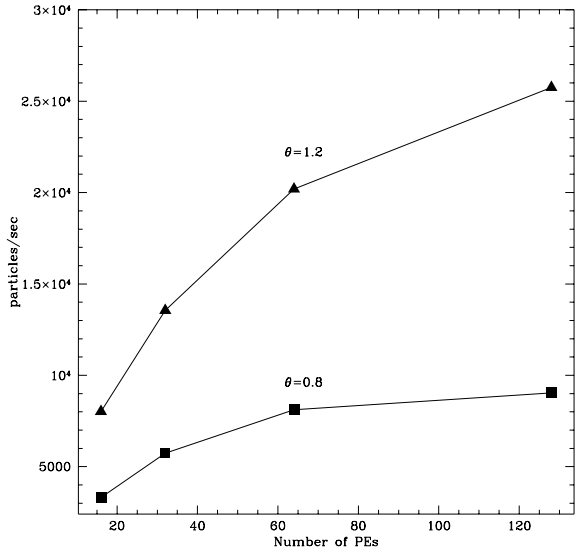,width=8cm}
\caption[h]{Code speed for clustered $10^6$ particles}
\end{figure}
The results obtained using the  WDSH-PTc code, at present,  give performances 
comparable to those obtained with different approaches like LET \cite{d96}, 
with the advantage of avoiding the LET and  an 
excessive demand for memory.

\section  {\bf Conclusions}

The study carried out on the CRAY T3D machines at the Cineca (Casalecchio di Reno, ITALY), allows us
 to  propose  
a criterion for the optimal data distribution of bodies and tree cells properties. A 
strategy for the 
{\it automatic}  Dynamic Load Balance has been described, which does not introduce a
significant overhead.
The results of this work will allow us to obtain, in the next future, a WDSH-PTc
 version 
for the CRAY T3E system, using  the HPF-CRAFT and the shmem library.
 The new version will include an enhanced grouping strategy 
and periodic boundary conditions, and will allow us to run large 
simulations with very high performances \cite{bapg97}. Using the CINECA Visualization Laboratory 
experience and resources, 
we plan also to develop an ad hoc visualization package for the scientific visualization of the 
simulation results

We would like to thank the CINECA for the supercomputing time grant, 
Mr. S. Bassini for the availability of the CINECA support and Dr. L. 
Calori, both from Cineca , and  Dr. F. Dur\'{\i} from Silicon Graphics
 for useful suggestions and comments. A. Pagliaro wish to tank the EPCC 
staff. 

{}
\clearpage

\clearpage

\end{document}